\documentclass[usenatbib]{mn2e}
\usepackage{graphicx}

\def\apj{\rm ApJ}
\def\apjl{\rm ApJL}
\def\apjs{\rm ApJS}

\def\mnras{\rm MNRAS}
\def\nat{\rm Nature}

\def\pasa{\rm PASA}
\def\aap{\rm AAP}
\def\araa{\rm ARA\&A}

\def\apss{\rm Ap\&SS}

\def\gax{\mathrel{\raise.3ex\hbox{$>$}\mkern-14mu\lower0.6ex\hbox{$\sim$}}}
\def\lax{\mathrel{\raise.3ex\hbox{$<$}\mkern-14mu\lower0.6ex\hbox{$\sim$}}}
\def\gtorder{\mathrel{\raise.3ex\hbox{$>$}\mkern-14mu
             \lower0.6ex\hbox{$\sim$}}}
\def\ltorder{\mathrel{\raise.3ex\hbox{$<$}\mkern-14mu
             \lower0.6ex\hbox{$\sim$}}}

\begin{document}

\title [The Progenitor of the Vela Pulsar]
   {The Progenitor of the Vela Pulsar}

\author[C.~S. Kochanek]{ 
    C.~S. Kochanek$^{1,2}$ 
    \\
  $^{1}$ Department of Astronomy, The Ohio State University, 140 West 18th Avenue, Columbus OH 43210 \\
  $^{2}$ Center for Cosmology and AstroParticle Physics, The Ohio State University,
    191 W. Woodruff Avenue, Columbus OH 43210 \\
   }

\maketitle

\begin{abstract}
With Gaia parallaxes it is possible to study the stellar populations associated
with individual Galactic supernova remnants (SNR) to estimate the mass of the
exploding star.  Here we analyze the luminous stars near the
Vela pulsar and SNR to find that its progenitor was probably ($\gtorder 90\%$)
low mass ($8.1$-$10.3M_\odot$).  The presence of the O star
$\gamma^2$~Vel a little over 100~pc from Vela is the primary ambiguity, as
including it in the analysis volume significantly increases the probability (to 5\%)
of higher mass ($>20M_\odot$) progenitors.  However, to be a high mass star
associated with $\gamma^2$~Vel's star cluster at birth, the progenitor would have to
a runaway star from an unbound binary with an unusually high velocity.
The primary impediment to analyzing large numbers of Galactic SNRs
in this manner is the lack of accurate distances.  This can likely be
solved by searching for absorption lines from the SNR in stars as a function
of distance, a method which yielded a distance to Vela in agreement with the
direct pulsar parallax. If Vela was a $10M_\odot$ supernova in an external
galaxy, the $50$~pc search region used in extragalactic studies would contain
only $\simeq 10\%$ of the stars formed in a 50~pc region around the progenitor
at birth and $\simeq 90\%$ of the stars in the search region would have been
born elsewhere.
\end{abstract}


\section{Introduction}

We would like to understand which massive stars explode as supernovae and the
nature of the resulting compact objects.  Modern theoretical models (e.g.,
\citealt{OConnor2011}, \citealt{Pejcha2015}, \citealt{Ertl2016},
\citealt{Sukhbold2016}, \citealt{Ghosh2021})
find a complex mapping between progenitor mass and explosion driven
by changes in core structure related to the balance between radiative
and convective carbon burning (\citealt{Sukhbold2018}, \citealt{Sukhbold2020}).
In these modern models, essentially all stars either explode and
produce neutron stars or fail to explode and produce black holes -- 
``fall back'' supernovae where the star explodes but a significant amount
of mass falls back onto the proto-neutron star to form a black hole are
extremely rare.\footnote{Despite this, fall back models for remnant masses such as
\cite{Fryer2012} remain in common use for binary population synthesis models
(e.g., \citealt{Belczynski2020}, 
\citealt{Breivik2020}, \citealt{Chawla2021}, \citealt{Eldridge2017}, \citealt{Riley2021}).}  
In the absence of fall back, lower mass ($\ltorder 10M_\odot$)
black holes are produced either from explosions of stars stripped by pre-supernova
mass loss/transfer or by the failed explosions of the more massive
red supergiants. These produce black holes with the mass of the helium 
core (\citealt{Kochanek2014}) because the \cite{Nadezhin1980} mechanism 
(also see, e.g., \citealt{Lovegrove2013}, 
\citealt{Fernandez2018}) ejects the weakly bound hydrogen envelope of
the supergiant.  None
of these surveys of outcomes are based on full {\it ab initio} core 
collapse simulations but instead use ``calibrated'' explosion models
to explore outcomes.  The outcomes in true core collapse simulations remain
an open problem because of the complexity of the physics and the need for high resolution
three-dimensional simulations (e.g., \citealt{Bollig2021},
\citealt{Burrows2020}, \citealt{Pan2021}).

To test these theoretical predictions, we need to observationally determine the mapping
between progenitors and outcomes. The cleanest approach for the stars which explode
is simply to measure the properties of the progenitor star, as first 
done for SN~1987A (\citealt{Gilmozzi1987}).  Since the stellar luminosity
is determined by the mass of the core, and the mass of the core is 
determined by the initial mass of the progenitor this is a fairly robust 
approach if there is sufficient data to well-determine the luminosity
(although the mass at the time of explosion cannot be well-constrained,
see \citealt{Farrell2020}).  The challenge is that this must be done in
distant ($\sim 1$-$10$~Mpc) galaxies and largely depends on the existence of
multi-band archival Hubble Space Telescope data for robust results.  
The progenitors of Type~IIP supernovae are red supergiants with an
upper mass limit that is consistent with the theoretical explosion 
studies (\citealt{Smartt2009a}, \citealt{Smartt2009b}, \citealt{Smartt2015}, although  
there are rebutted (e.g., \citealt{Kochanek2012}, \citealt{Kochanek2020}, \citealt{Beasor2020}) 
counterarguments (e.g., \citealt{Walmswell2012}, \citealt{Groh2013}, 
\citealt{Davies2020}).  Less is
known about the progenitors of Type~Ibc supernovae because they
are generally undetected, which likely implies that in most cases
the envelopes are stripped by binary processes (e.g., \citealt{Eldridge2013},
\citealt{Folatelli2016}, \citealt{Johnson2017}, \citealt{Kilpatrick2021}).
Unfortunately, the nature of the compact remnant formed in these systems
will likely always be unknown.

Failed supernovae can be found by searching for stars which disappear
independent of the nature of any associated transient (\citealt{Kochanek2008}).
A search for failed supernovae using the Large Binocular Telescope 
(\citealt{Gerke2015}, \citealt{Adams2017b}, \citealt{Neustadt2021})
has identified one excellent candidate (\citealt{Gerke2015},
\citealt{Adams2017a}, \citealt{Basinger2020}) and a second,
weaker candidate (\citealt{Neustadt2021}).  As expected, the
progenitor of the strong candidate was a massive RSG, and
finding one candidate implies a fraction of core collapses
leading to failed supernovae consistent with the current theoretical
predictions.

The primary alternative to searching for individual progenitors is to 
use the local stellar population to infer the probable mass of the
progenitor.  This has been done for supernova remnants in the 
Magellanic Clouds (\citealt{Badenes2009}, \citealt{Auchettl2019})
and in nearby galaxies (e.g., \citealt{Jennings2012}, \citealt{Jennings2014},
\citealt{Diaz2018}, \citealt{Williams2018},
\citealt{Williams2019}, \citealt{Diaz2021}, \citealt{Koplitz2021}).
Except for \cite{Williams2018} and \cite{Diaz2021}, which examine
stellar populations near known supernovae, these are
studies of the stellar populations near supernova remnants. The primary
advantage of this method is that it can be applied to large numbers
of supernova remnants or historical supernovae compared to the numbers
of directly observed progenitors.  For the Magellanic Clouds, several
of the remnants are associated with neutron stars (N49, N158A and possibly
SN~1987A in the LMC, and 1KT6 and 1E~0102.2$-$7219 in the SMC,
see \citealt{Badenes2009}, \citealt{Auchettl2019}), so the
outcome of the explosion is also known. 

There are
also several disadvantages. First, unless the local stellar population has 
a single well-defined starburst, the method does not provide individual
well-constrained masses and so is best suited for making statistical
models of the progenitor distribution.  Second, the nature of the supernova is 
unknown for SNRs, so, for example, one cannot separately investigate the progenitors
of Type~IIP and Type~Ibc supernovae.  
No SNR studies can directly address the deficit
of more massive RSG progenitors to Type~IIP supernovae because the
supernova types are unknown.  
Third, the life time and detectability of the
SNR has some dependence on the nature of the explosion (e.g., \citealt{Sarbadhicary2017},
\citealt{Jacovich2021})
which will introduce some bias into the
results.  Most, but not all, of these studies have supported a deficit
of higher mass SN progenitors.  
The progenitor mass distribution models 
used to date in these studies are simple functional forms that do not resemble current 
theoretical expectations, so it is difficult to evaluate the degree to which 
they agree or disagree with these expectations.
  
With the advent of Gaia (\citealt{Gaia2016}, \citealt{Gaia2021}) it is now
possible to apply this second method to Galactic supernova remnants if the distance
to the remnant is known and the extinction is not severe.  The present
number of such systems is small, primarily because so few supernova 
remnants have well-constrained distances.  As with extragalactic SNRs,
the supernova type will generally be unknown, but, unlike extragalactic SNRs, the
compact object outcome of the explosion frequently is known.  Because the Galaxy is so well-surveyed
from the mid-IR into the ultraviolet, it will generally be possible to characterize
the individual stars extremely well, with well-determined individual stellar luminosities,
temperatures and extinctions.  In many cases, the brighter stars will also have
spectroscopic classifications.   Here we demonstrate this for the Vela 
pulsar (PSR~J0835$-$4510, PSR~B0833$-$45, \citealt{Large1968}), which
is so nearby (280~pc, \citealt{Dodson2003}) that many of the 
massive stars near it are visible to the naked eye.  We describe the selection
of the stars and spectral energy distribution (SED) models of the more
luminous stars in \S2.  We analyze these stars to estimate the likely mass of Vela's progenitor
in \S3, and we discuss the results and future prospects in \S4.

\section{Selecting the Stars}

We select the stars from the Gaia EDR3 catalog (\citealt{Gaia2016}, 
\citealt{Gaia2021}), requiring them to have parallaxes,
 proper motions and all three Gaia magnitudes.
We apply no restriction on the RUWE statistic for the quality 
of the parallax.
We adopt the position (J2000 08:35:20.61149$\pm$0.00002, $-$45:10:34.8751$\pm$0.0003), 
parallax ($\varpi = 3.5 \pm 0.2$~mas) and proper motions
($\mu_\alpha = -49.68\pm0.06$ and $\mu_\delta = 29.9 \pm 0.1$~mas/year)
of the pulsar from \cite{Dodson2003}. The parallax is sufficiently accurate that it is essentially
unaffected by Lutz-Kelker bias (see \citealt{Verbiest2012}).
The spin down age of Vela is $11.3$~kyr, although \cite{Lyne1996} argue
that the braking index implies a larger age.
The proper motions combined with the offset from the center
of the remnant imply an age of $18000 \pm 9000$~years based on the
$25\pm 5$~arcmin offset of the pulsar from the center of the
remnant (\citealt{Aschenbach1995}).  Unfortunately, \cite{Aschenbach1995}
do not report their estimated position for the center, but if  we adopt an 
age of $T=20000$ years we can estimate the position of the explosion
from the position shift of ($+17$,$-10$)~arcmin.  The remnant itself 
has a diameter of approximately $8.3$~degrees, so the shift is small
compared to the scale of the remnant and the size of our search
region.  
We select the stars in a region centered on the shifted equatorial position of
($128.84^\circ$, $-45.18^\circ$) where we have reduced the coordinate precision
due to the uncertain age.  

The next question is the size of the region to extract around Vela
to obtain a representative stellar population.  The studies of the
regions around supernovae and SNRs in nearby galaxies have almost
uniformly used a region 50~pc in radius based on the argument that
stars form in compact clusters ($\sim 1$~pc, \citealt{Lada2003}),
drift apart at low relative velocities ($\sim$~km/s), and that
stars which explode after their parent binary was disrupted by a
supernova have only modest velocities (\citealt{Eldridge2011},
\citealt{Renzo2019}).  With Gaia, we can simply consider this 
question empirically.  In this section we consider the three
dimensional geometry we use for Vela, and in \S4 we consider
the two-dimensional projected geometry of extragalactic analyses.

We started by extracting all stars within $R=250$~pc of Vela.
If the pulsar is at a distance of $D=286$~pc, a 
sphere of radius $R$ subtends a maximum angle of 
$\theta = \sin^{-1} R/D \simeq 62$~deg relative to its center.
We select stars with $\theta < 62$~deg, 
$1.87 < \varpi < 28.6$~mas, which corresponds
to $\pm 250$~pc around the pulsar distance, and $G < 9.5$~mag. 
This will include all stars with $M_G \ltorder 0$~mag since the
extinctions are small. At this magnitude limit we are including
all $M \gtorder 5M_\odot$ stars and the more luminous, evolved,
lower mass stars.  The initial search yields $\sim 55,000$ stars, 
of which $\sim 37000$ are within $R=250$~pc of the pulsar, and 
$3160$ are brighter than $M_G< 0$~mag.

\begin{figure}
\includegraphics[width=0.50\textwidth]{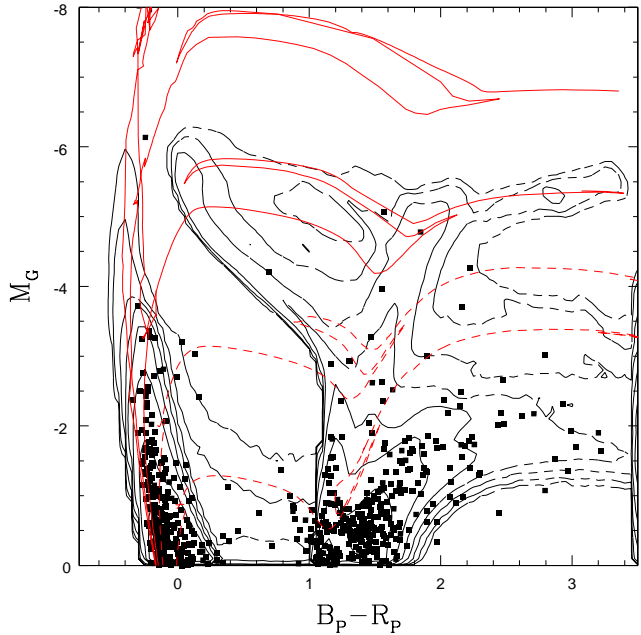}
\caption{
  Color-magnitude diagram without extinction corrections for the stars with 
  $M_G>0$~mag and within $R<125$~pc of Vela.  The background contours are the maximum
  likelihood model for the stellar density including extinction with solid (dashed)
  contours at higher (lower) densities that are spaced by factors of $3$.
  The red curves are Solar metallicity {\tt PARSEC} 
  isochrones with ages of $10^{8.5}$ (bottom, dashed), $10^{8.0}$ (dashed)
  $10^{7.5}$ (solid), $10^{7.0}$ (solid) and $10^{6.6}$ (top, solid) years.
  The maximum mass for the $10^{7.5}$~year isochrone is $9.1M_\odot$, so
  only the solid isochrones still have stars which will explode as 
  supernovae.  The isochrones are shifted by the mean extinction of
  $E(B-V) = 0.057$~mag.
  }
\label{fig:cmd}
\end{figure}

The bright magnitude limit of Gaia at $G \simeq 3$~mag can be a 
problem because the most luminous stars may not be 
present in the Gaia catalog.  We searched the Hipparcos
(\citealt{Perryman1997}, using the updated 
astrometric solution from \citealt{vanLeeuwen2007}) and
the Bright Star (\citealt{Hoffleit1995}) catalogs for any
such stars in the search volume, finding nine Hipparcos
stars: 
HIP~30324 ($\beta$~CMa)
HIP~33579 ($\epsilon$~CMa),
HIP~39953 ($\gamma^2$~Vel),
HIP~41037 ($\epsilon$~Car) 
HIP~44816 ($\lambda$~Vel),
HIP~66657 ($\epsilon$~Cen)
HIP~68702 ($\zeta$~Cen), and
HIP~81173 ($\alpha$~TrA).  
Since $\gamma^2$~Vel is in a wide binary (\citealt{Tokovinin2018})
with $\gamma^1$~Vel (HD~68243), we use the Gaia parallax
of $\gamma^1$~Vel for $\gamma^2$~Vel.  For the other 
stars we adopt the updated Hipparcos parallaxes.
Including the bright Hipparcos stars, there are $N=3169$ stars 
with $M_G < 0$~mag within $R=250$~pc of Vela.  
Figure~\ref{fig:cmd} shows the distribution of the $R<125$~pc
stars in absolute 
magnitude $M_G$ and color $B_P-R_P$ (with no extinction corrections) 
as compared to Solar metallicity
{\tt PARSEC} (\citealt{Bressan2012}, \citealt{Marigo2013}, \citealt{Pastorelli2020}) 
isochrones with ages of $10^{6.6}$, $10^{7.0}$, $10^{7.5}$,
$10^{8.0}$ and $10^{8.5}$~years.  For the bright Hipparcos stars we
used the SED fits described below to synthesize the Gaia magnitudes. 

To explore optimizing the region from which we select stars, we
only want to consider the younger, massive stars.
The main contaminants are lower mass red giants, which we can
largely remove by considering only
stars with $M_G < 0$ for $B_P-R_P < 0.6$, $M_G < -2.0$
for $0.6 < B_P-R_P < 1.8$, and $M_G < -3.5$ for
$B_P-R_P > 1.8$.  
Only a small fraction of the stars have Gaia DR2
radial velocities, so we assigned a random line of sight velocity
with a dispersion of $3.0$~km/s based on the velocity dispersion
derived from the proper motions of the stars within 50~pc of 
Vela (see below).

In our present analysis we are selecting the stars in three
dimensions, so we want to explore the effect of changing the
radius of the selection sphere.  We can estimate the 
completeness and contamination of the selection by examining
the relative positions of the stars today and in the past.  
We start by taking the stars near the
pulsar and determining where they were $10^7$~years and 
$10^{7.5}$ years in the past -- these roughly correspond
to the lifetimes of $20M_\odot$ and $10M_\odot$ stars,
respectively.  Because the pulsar received a kick, we
cannot do the same for the past location of the progenitor
star.  For simplicity, we assume that the progenitor had
the mean velocity of the stars.  The completeness will
be lower if we include a random motion for the progenitor.

Following the extragalactic studies, we first considered a
50~pc sphere.  This sphere around Vela today contains 19
of these young, high mass stars.  After subtracting their
median tangential velocities, they have a two-dimensional (2D)
velocity dispersion of 4.2~km/s defined by half the width
of the velocity range encompassing 68\% of the stars. 
As we expand the radius of the sphere, the number of stars
rises rapidly to 152, 411, 757 and 1160 within 100, 150, 200
and 250~pc, and the 2D velocity dispersion rises slowly to 
$5.4$, $6.7$, $6.7$ and $7.9$~km/s, respectively.  We 
used the 4.2~km/s dispersion derived from the proper motions
to set the $4.2/\sqrt{2}=3.0$~km/s dispersion of the randomly
assigned line of sight velocity.  The three dimensional velocity
dispersion is then $5.1$~km/s, so the typical typical massive 
has a random motion of $\simeq 50$~pc in the lifetime of a 
$\sim 20M_\odot$ star ($\sim 10^7$~years) and
$\simeq 150$~pc in the lifetime of a $\sim 10M_\odot$ 
star ($\sim 10^{7.5}$~years).

If we place a sphere at the median position of these stars 
$10^7$ years ago, we can estimate the completeness and
contamination.  The completeness is the fraction of stars
inside a sphere of some radius today that are inside a
sphere of some other radius centered at the median position
of the stars in the past.   The contamination is the fraction
of the stars in the ``past sphere'' that were not in the
``present sphere''.  For a 50~pc sphere today, 4, 12,
16 and 18 of the 19 stars are in a past sphere of radius
$50$, $100$, $150$, and $200$~pc, along with 10, 61, 179
and 356 stars which were not in the 50~pc sphere today.
So the completenesses are 21\%, 63\%, 84\% and 95\%, while
the contamination rates are 71\%, 84\%, 92\% and 95\%. 
Basically, since
the typical random motion corresponds to moving $50$~pc
in $10^7$~years, a sphere of comparable size must suffer
from poor completeness and significant contamination.
If we increase the size of the present sphere to 
$100$~pc, the completenesses in past spheres of radius
$100$, $150$ and $200$~pc are 36\%, 68\% and 84\%,
respectively, and the contamination rates are 21\%,
47\% and 67\%.  Finally, if we use $150$~pc, the
completeness/contamination are 40\%/14\%  and 70\%/25\%
for past spheres of radius $150$ and $200$~pc.  
This obviously gets much worse if we use the $10^{7.5}$~year
lifetime of a $\simeq 10M_\odot$ star.  For a $50$~pc sphere
today, and 50, 100, 150 and 200~pc spheres $10^{7.5}$~years
ago, the completeness/contamination rates are 
5\%/75\%, 16\%/79\%, 26\%/88\% and 47\%/89\%, respectively
(3\%/61\%, 9\%/61\%, and 22\%/58\% for $100$~pc today;
7\%/15\% and 15\%/25\% for $150$~pc today).  

Completeness and contamination are, or course, a trade off -- there is
no right answer.  We will present all of our results for
a radius of $125$~pc.  We were originally going to use $100$~pc
but the most massive star near Vela, $\gamma^2$~Vel, lies just
outside a $100$~pc sphere, so we increased the radius to $125$~pc
to double the volume and include $\gamma^2$~Vel.  For the final
results on the probable age and mass of Vela's progenitor, we
will present the results for a range of radii.

We fit the spectral energy distributions (SEDs) 
of the more luminous stars in the $125$~pc sphere to estimate 
their luminosity, temperature and extinctions.  
We also fit the bright Hipparcos stars to synthesize
their Gaia magnitudes.
The SED fits are moderately labor intensive in terms of collating the data,
so we did not model all the lower luminosity stars. 
We initially selected all stars with
$M_G < -2$ ($M_G < -3.5$) for $B_P-R_P < 1.8$ ($>1.8$), where the
higher luminosity limit for redder stars eliminates lower mass
red giants.  While the SEDs
showed no need to include circumstellar dust, we used the same methods
as in \cite{Adams2017a}, running {\tt DUSTY} (\citealt{Elitzur2001}) 
inside a Markov Chain Monte
Carlo (MCMC) driver to both optimize the fits and then estimate
the uncertainties.  We used \cite{Castelli2003} model atmospheres
for all but the coolest stars where we used {\tt MARCS} 
(\citealt{Gustafsson2008}) model atmospheres.
We used near-IR data from 2MASS (\citealt{Cutri2003}).
The optical data primarily came from \cite{Johnson1966} and \cite{Tonry2018},
supplemented by \cite{Cousins1971}, \cite{Ducati2002}, \cite{Mermillod1977},
\cite{Morel1978}, NOMAD (\citealt{Zacharias2005}), and \cite{Neckel1980}.
Almost all of the hot stars had UV data extending to $\sim 1500$~\AA\
from either \cite{Thompson1978} or \cite{Wesselius1982}.

We used temperature priors based on the reported
spectral types and weak extinction priors.
The temperature prior widths were roughly one spectral
type (so B1 to B3 for a B2 star) and $\pm 0.1$~mag for $E(B-V)$.
For extinction priors, we use the three dimensional {\tt combined19 mwdust} 
models (\citealt{Bovy2016}), which combine the \cite{Drimmel2003}, \cite{Marshall2006}
and \cite{Green2019} models to provide estimates for any sky position. We
extracted the V band extinction, and then used an $R_V=3.1$ extinction law
to convert the V band extinction to those for the $G$, $B_P$ and $R_P$ bands.
For most stars, the SEDs could only modestly improve the spectral
temperature estimates but strongly constrained the extinction. 
The agreement with the {\tt mwdust} estimates was generally good,
except when the {\tt mwdust} estimate was high ($E(B-V) \gtorder 0.1$).
In these cases, the SED models generally required much less dust.
With this caveat, we will use the {\tt mwdust} extinctions to model
the effects of extinction for the full sample.

Table~\ref{tab:stars} gives the goodness of fits, temperature,
luminosity, mass, age, separation from estimated explosion center
and some comments for the stars with $L_*>10^{3.5}L_\odot$.  
The individual masses and
ages are simply the range of PARSEC isochrone ages and masses 
where the luminosity and temperature are within twice the
estimated uncertainties of the fitted values with a minimum
uncertainty of $0.02$~dex.  The ages and
masses are strongly correlated - the maximum masses correspond
to the minimum ages and {\it vice versa}. 
Except for $\gamma^2$~Vel, they are all less massive than $\simeq 15M_\odot$ and
older than $\simeq 10^7$~years even for the upper (lower) limits on the masses (ages).

While many of the stars are binaries (see Table~\ref{tab:stars}), the
extra light from the secondary seems to have little
consequence for the SED fits.  This is not very surprising because
B stars are still in the regime where the mass-luminosity relation
is fairly steep and a modestly lower mass companion will not greatly
perturb the SED.   For example, consider the most luminous star, 
$\gamma^2$~Vel, which is an O star plus Wolf-Rayet (WR) star binary.
$\gamma^2$~Vel is an interferometrically
resolved double lined spectroscopic binary with present day masses of $(28.5\pm1.1) M_\odot$  
and $(9.0\pm0.6)M_\odot$ for the O star and the companion WR
star, respectively (\citealt{North2007}).  While our SED fit finds a higher
total luminosity, most of the difference is due to both the Gaia EDR3
distance (where we used the distance to the wide binary companion $\gamma^1$~Vel)
and our extinction estimate being larger than those used by \cite{North2007}.  
Still, our rough individual mass estimate of $25.3$-$27.3M_\odot$ agrees
reasonably well with their dynamical measurement.  \cite{North2007} estimate that only
$22\%$ of the V-band light comes from the WR star, so the SED fit itself
is not strongly biased by the WR companion.  

\begin{table*}
  \centering
  \caption{Luminous Stars Near the Vela Pulsar}
  \begin{tabular}{lrccccrl}
  \hline
  \multicolumn{1}{c}{Star} &
  \multicolumn{1}{c}{$\chi^2/N_{dof}$} &
  \multicolumn{1}{c}{$\log(T_*)$} &
  \multicolumn{1}{c}{$\log(L_*)$} &
  \multicolumn{1}{c}{$M_*$} &
  \multicolumn{1}{c}{$\log t$} &
  \multicolumn{1}{c}{Sep} &
  \multicolumn{1}{c}{Comments}
   \\
  &
  &
  \multicolumn{1}{c}{(K)} &
  \multicolumn{1}{c}{($L_\odot$)} &
  \multicolumn{1}{c}{($M_\odot$)} &
  \multicolumn{1}{c}{(yr)} &
  \multicolumn{1}{c}{(pc)} \\
 
  \hline
      HD~68273 &$ 0.79 $ &$  4.622 \pm  0.020 $ &$  6.029 \pm  0.070 $ &$ 25.3 $-$ 27.3 $ &$ 6.61 $-$ 6.63 $ & 102 & $\gamma^2$~Vel, O7.5III+WC8 binary \\ 
    HD~63462 &$ 1.82 $ &$  4.383 \pm  0.032 $ &$  4.347 \pm  0.094 $ &$  9.8 $-$ 13.9 $ &$ 6.96 $-$ 7.40 $ & 124 & O~Pup, B1IVe \\ 
    HD~68243 &$ 1.00 $ &$  4.358 \pm  0.033 $ &$  4.319 \pm  0.084 $ &$  9.5 $-$ 13.2 $ &$ 7.07 $-$ 7.43 $ & 102 & $\gamma^1$~Vel, B2III \\ 
    HD~65818 &$ 1.74 $ &$  4.390 \pm  0.029 $ &$  4.297 \pm  0.075 $ &$  9.9 $-$ 13.4 $ &$ 6.92 $-$ 7.39 $ & 92 & V~Pup, B1Vp+B2 ecl. binary \\ 
    HD~74575 &$ 0.27 $ &$  4.345 \pm  0.018 $ &$  4.149 \pm  0.053 $ &$  8.9 $-$ 11.3 $ &$ 7.19 $-$ 7.49 $ & 70 & $\alpha$~Pyx, B1.5III \\ 
     HR~3307 &$ 2.29 $ &$  3.624 \pm  0.012 $ &$  4.137 \pm  0.029 $ &$  7.9 $-$ 13.1 $ &$ 7.19 $-$ 7.62 $ & 116 & $\epsilon$~Car, K3III+B2 binary \\ 
    HD~57150 &$ 1.03 $ &$  4.325 \pm  0.040 $ &$  4.076 \pm  0.121 $ &$  7.7 $-$ 11.5 $ &$ 7.10 $-$ 7.61 $ & 83 & $\upsilon^1$~Pup, B2V+B3IV binary \\ 
    HD~78647 &$ 2.03 $ &$  3.618 \pm  0.010 $ &$  4.050 \pm  0.044 $ &$  2.0 $-$ 12.1 $ &$ 7.25 $-$ 8.73 $ & 121 & $\lambda$~Vel, K4I \\ 
    HD~63032 &$ 1.53 $ &$  3.619 \pm  0.010 $ &$  4.008 \pm  0.036 $ &$  2.0 $-$ 11.7 $ &$ 7.27 $-$ 8.74 $ & 84 & c~Pup, K4III \\ 
    HD~65551 &$ 1.61 $ &$  4.407 \pm  0.014 $ &$  3.988 \pm  0.034 $ &$  9.8 $-$ 11.9 $ &$< 7.27$          & 51 & N~Pup, B2II/IV \\ 
    HD~83058 &$ 0.57 $ &$  4.345 \pm  0.035 $ &$  3.940 \pm  0.098 $ &$  7.6 $-$ 10.8 $ &$ 6.84 $-$ 7.62 $ & 90 & L~Vel, B2IV \\ 
    HD~56139 &$ 2.75 $ &$  4.246 \pm  0.028 $ &$  3.931 \pm  0.087 $ &$  7.3 $-$  9.3 $ &$ 7.42 $-$ 7.66 $ & 120 & $\omega$~CMa, B2.5Ve \\ 
    HD~51799 &$ 2.58 $ &$  3.597 \pm  0.008 $ &$  3.862 \pm  0.034 $ &$  1.9 $-$ 10.0 $ &$ 7.38 $-$ 8.75 $ & 122 & M1III \\ 
    HD~63465 &$ 0.93 $ &$  4.261 \pm  0.032 $ &$  3.843 \pm  0.108 $ &$  6.6 $-$  9.3 $ &$ 7.41 $-$ 7.75 $ & 105 & B2IV/V \\ 
    HD~68324 &$ 0.31 $ &$  4.353 \pm  0.028 $ &$  3.835 \pm  0.075 $ &$  8.2 $-$ 10.3 $ &$ 6.62 $-$ 7.47 $ & 65 & IS~Vel, B2V \\ 
    HD~64740 &$ 0.84 $ &$  4.369 \pm  0.024 $ &$  3.800 \pm  0.059 $ &$  8.4 $-$ 10.4 $ &$< 7.42$          & 56 & B2V \\ 
    HD~73155 &$ 1.78 $ &$  3.671 \pm  0.011 $ &$  3.799 \pm  0.040 $ &$  6.6 $-$  9.5 $ &$ 7.43 $-$ 7.78 $ & 90 & C~Vel, K1.5I \\ 
    HD~68217 &$ 0.34 $ &$  4.324 \pm  0.034 $ &$  3.772 \pm  0.094 $ &$  6.9 $-$  9.7 $ &$ 6.79 $-$ 7.70 $ & 48 & B2IV \\ 
    HD~64503 &$ 0.44 $ &$  4.307 \pm  0.029 $ &$  3.727 \pm  0.081 $ &$  6.8 $-$  9.0 $ &$ 7.17 $-$ 7.72 $ & 64 & b~Pup, B2V binary \\ 
    HD~65575 &$ 0.73 $ &$  4.232 \pm  0.029 $ &$  3.661 \pm  0.099 $ &$  6.0 $-$  8.1 $ &$ 7.53 $-$ 7.84 $ & 117 & $\chi$~Car, B3IV \\ 
    HD~57219 &$ 0.25 $ &$  4.263 \pm  0.030 $ &$  3.643 \pm  0.087 $ &$  6.1 $-$  8.2 $ &$ 7.42 $-$ 7.83 $ & 102 & $\upsilon^2$~Pup, B3Ve \\ 
    HD~89388 &$ 1.63 $ &$  3.639 \pm  0.008 $ &$  3.642 \pm  0.035 $ &$  5.2 $-$  8.7 $ &$ 7.50 $-$ 8.03 $ & 112 & q~Car, K2.5II \\ 
    HD~85123 &$ 0.75 $ &$  3.879 \pm  0.018 $ &$  3.635 \pm  0.041 $ &$  6.0 $-$  7.4 $ &$ 7.64 $-$ 7.86 $ & 108 & $\upsilon$~Car, A9 \\ 
    HD~54893 &$ 0.32 $ &$  4.282 \pm  0.028 $ &$  3.602 \pm  0.082 $ &$  6.2 $-$  8.2 $ &$ 7.30 $-$ 7.80 $ & 83 & A~Pup, B3IV/V \\ 
    HD~63949 &$ 0.58 $ &$  4.286 \pm  0.029 $ &$  3.563 \pm  0.091 $ &$  6.2 $-$  8.1 $ &$ 7.21 $-$ 7.81 $ & 112 & QS~Pup, B2III $\beta$~Ceph \\ 
    HD~79275 &$ 0.91 $ &$  4.320 \pm  0.025 $ &$  3.509 \pm  0.073 $ &$  6.8 $-$  8.5 $ &$< 7.60$          & 63 & B2III/IV \\ 
    HD~56779 &$ 0.61 $ &$  4.247 \pm  0.024 $ &$  3.490 \pm  0.074 $ &$  5.7 $-$  7.3 $ &$ 7.53 $-$ 7.89 $ & 85 & B3V \\ 
    HD~69144 &$ 0.79 $ &$  4.195 \pm  0.025 $ &$  3.487 \pm  0.084 $ &$  5.4 $-$  7.1 $ &$ 7.66 $-$ 7.94 $ & 82 & NO~Vel, B3III ecl. binary \\ 
    HD~84816 &$ 1.06 $ &$  4.268 \pm  0.026 $ &$  3.465 \pm  0.082 $ &$  5.9 $-$  7.5 $ &$ 7.36 $-$ 7.86 $ & 82 & B2V \\ 
    HD~85622 &$ 2.08 $ &$  3.697 \pm  0.012 $ &$  3.438 \pm  0.046 $ &$  5.3 $-$  7.2 $ &$ 7.67 $-$ 8.01 $ & 66 & m~Vel, G6IIa binary \\ 
    HD~63922 &$ 0.42 $ &$  4.152 \pm  0.020 $ &$  3.429 \pm  0.053 $ &$  5.6 $-$  6.4 $ &$ 7.77 $-$ 7.92 $ & 40 & P~Pup, B7/8 \\ 
    HD~61641 &$ 0.78 $ &$  4.246 \pm  0.023 $ &$  3.402 \pm  0.073 $ &$  5.6 $-$  7.0 $ &$ 7.51 $-$ 7.92 $ & 106 & e~Pup, B3III \\ 
    HD~59550 &$ 0.89 $ &$  4.269 \pm  0.020 $ &$  3.401 \pm  0.060 $ &$  6.2 $-$  7.2 $ &$ 7.36 $-$ 7.74 $ & 118 & B2IV \\ 
    HD~71510 &$ 0.41 $ &$  4.298 \pm  0.047 $ &$  3.394 \pm  0.132 $ &$  5.6 $-$  8.6 $ &$< 7.91$          & 62 & B3IV, Be star \\ 
    HD~68895 &$ 0.70 $ &$  4.203 \pm  0.026 $ &$  3.295 \pm  0.086 $ &$  5.0 $-$  6.5 $ &$ 7.67 $-$ 8.04 $ & 104 & B5V binary \\ 
    HD~76566 &$ 1.02 $ &$  4.244 \pm  0.025 $ &$  3.251 \pm  0.080 $ &$  5.5 $-$  6.6 $ &$ 7.36 $-$ 7.84 $ & 89 & B3IV \\ 
    HD~75710 &$ 1.67 $ &$  3.959 \pm  0.005 $ &$  3.251 \pm  0.025 $ &$  5.0 $-$  5.6 $ &$ 7.91 $-$ 8.03 $ & 101 & g~Vel, A2III \\ 
    HD~82150 &$ 4.19 $ &$  3.630 \pm  0.008 $ &$  3.173 \pm  0.042 $ &$  2.5 $-$  6.2 $ &$ 7.81 $-$ 8.93 $ & 92 & $\epsilon$~Ant, K3III \\ 
    HD~91504 &$ 0.28 $ &$  3.675 \pm  0.011 $ &$  3.156 \pm  0.029 $ &$  4.5 $-$  6.2 $ &$ 7.81 $-$ 8.18 $ & 121 & t~Vel, K1/2III \\ 
    HD~72227 &$ 0.77 $ &$  3.655 \pm  0.012 $ &$  3.129 \pm  0.038 $ &$  3.4 $-$  6.1 $ &$ 7.83 $-$ 8.53 $ & 85 & K3III \\ 
    HD~92449 &$ 0.49 $ &$  3.697 \pm  0.013 $ &$  3.105 \pm  0.049 $ &$  4.5 $-$  5.9 $ &$ 7.86 $-$ 8.18 $ & 110 & x~Vel, G5II \\ 
    HD~50235 &$ 0.16 $ &$  3.646 \pm  0.010 $ &$  3.087 \pm  0.029 $ &$  2.5 $-$  5.9 $ &$ 7.86 $-$ 8.91 $ & 111 & K2/3III \\ 
    HD~85355 &$ 0.95 $ &$  4.100 \pm  0.013 $ &$  3.057 \pm  0.040 $ &$  4.5 $-$  5.1 $ &$ 7.99 $-$ 8.14 $ & 66 & n~Vel, B7III \\ 
    HD~75630 &$ 1.42 $ &$  3.958 \pm  0.006 $ &$  3.027 \pm  0.029 $ &$  4.4 $-$  4.8 $ &$ 8.06 $-$ 8.17 $ & 94 & h~Vel, A2/EIV \\ 
    HD~49689 &$ 0.48 $ &$  3.659 \pm  0.015 $ &$  3.026 \pm  0.051 $ &$  3.2 $-$  5.7 $ &$ 7.88 $-$ 8.59 $ & 94 & K3III \\ 
    HD~65699 &$ 0.22 $ &$  3.694 \pm  0.013 $ &$  2.933 \pm  0.043 $ &$  4.2 $-$  5.3 $ &$ 7.96 $-$ 8.27 $ & 112 & 12~Pup, G9IIa pec \\ 
    HD~68512 &$ 0.50 $ &$  3.668 \pm  0.013 $ &$  2.922 \pm  0.044 $ &$  3.2 $-$  5.3 $ &$ 7.96 $-$ 8.60 $ & 80 & K3III \\ 
    HD~85656 &$ 0.55 $ &$  3.678 \pm  0.012 $ &$  2.863 \pm  0.029 $ &$  3.4 $-$  5.1 $ &$ 8.00 $-$ 8.53 $ & 101 & K1III \\ 
    HD~76304 &$ 0.90 $ &$  3.692 \pm  0.009 $ &$  2.708 \pm  0.017 $ &$  3.4 $-$  4.7 $ &$ 8.10 $-$ 8.53 $ & 119 & K0II/III \\ 
    HD~80126 &$ 0.23 $ &$  3.726 \pm  0.012 $ &$  2.696 \pm  0.040 $ &$  3.8 $-$  4.5 $ &$ 8.13 $-$ 8.37 $ & 118 & G8II \\ 

  \hline
  \end{tabular}
  \label{tab:stars}
\end{table*}

\begin{figure}
\includegraphics[width=0.50\textwidth]{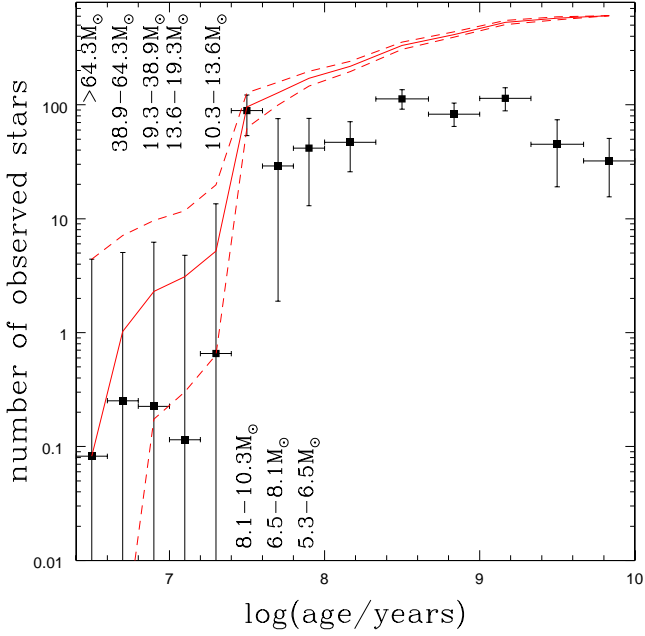}
\caption{
  Age distribution $N_i F_i$ of the modeled stars. The points show the 
  median number of stars associated with each age bin and
  the 90\% confidence range.  The horizontal error bars span
  the bin widths, and the mass range corresponding to the
  more massive age bins is listed.  The solid red curves shows
  the median integral distribution and the dashed curves
  show its 90\% confidence range.  All uncertainties are
  highly correlated.  The integral distribution converges
  exactly to the number of stars because we are determining
  how to distribute the modeled stars over the age bins.
  }
\label{fig:age}
\end{figure}

\begin{figure}
\includegraphics[width=0.50\textwidth]{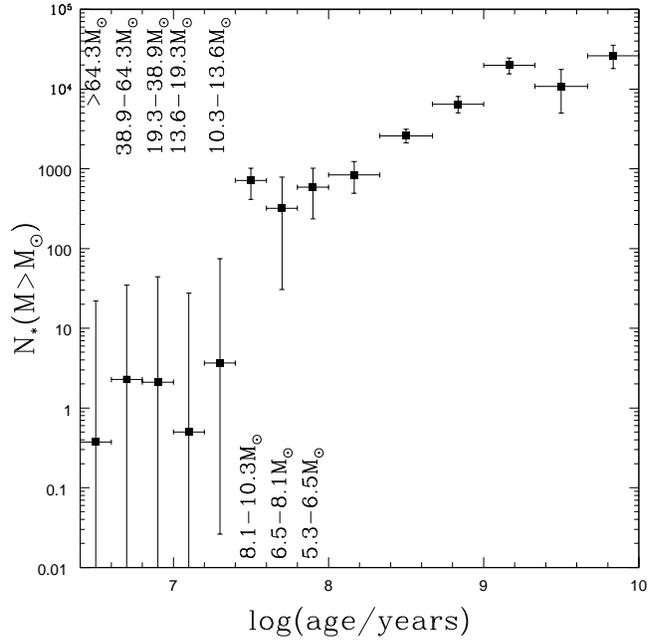}
\caption{
  Inferred numbers $N_i$ of $M>M_{min}=M_\odot$ stars formed for
  each age bin.  This corresponds to the number of modeled stars
  shown in Figure~\protect\ref{fig:age} divided by the fraction $F_i$ 
  of $M>M_{min}$ stars which have $M_G>0$~mag.  The points show the
  median number of $M>M_{min}$ stars associated with each age bin and
  the 90\% confidence range.  The horizontal error bars span
  the bin widths, and the mass range corresponding to the
  more massive age bins is listed.  All uncertainties are
  highly correlated.  
  }
\label{fig:age2}
\end{figure}

\begin{figure}
\includegraphics[width=0.50\textwidth]{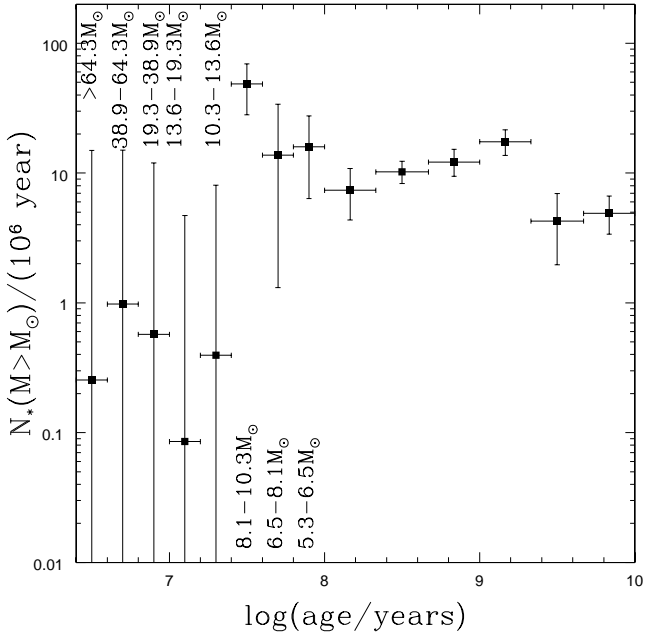}
\caption{
  Rate of forming $M>M_{min}=M_\odot$ stars per million years
  for each age bin.  This corresponds to the number of $M>M_{min}$
  stars $N_i$ shown in Figure~\protect\ref{fig:age2} divided by the
  temporal bin width $\Delta t_i$.  The points show the
  median star formation rate associated with each age bin and
  the 90\% confidence range.  The horizontal error bars span
  the bin widths, and the mass range corresponding to the
  more massive age bins is listed.  All uncertainties are
  highly correlated.  
  }
\label{fig:sfr}
\end{figure}

\begin{figure}
\includegraphics[width=0.50\textwidth]{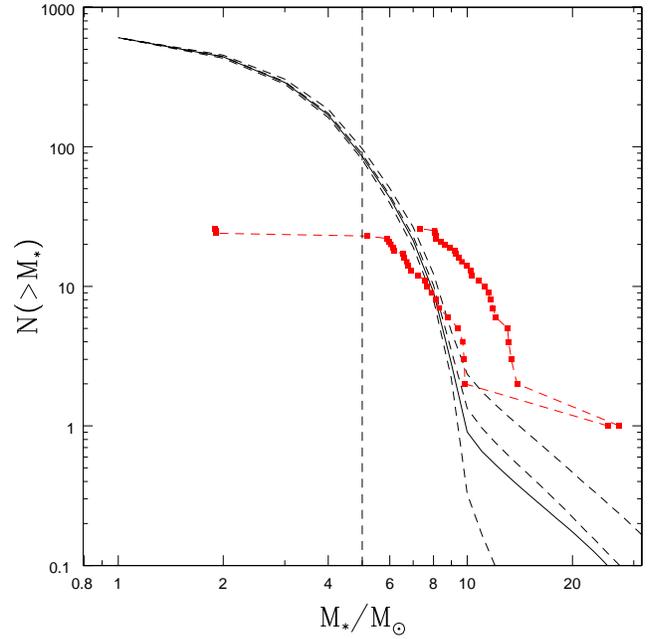}
\caption{
  Mass distribution of the $M_G>0$~mag stars.  The black solid shows
  the maximum likelihood model and the
  dashed curves show the mean and 90\% confidence range for the
  integral distributions from the MCMC chains.  This 
  distribution should be nearly complete for masses above 
  $5M_\odot$ (the vertical dashed line) and increasingly
  incomplete for lower masses.   The red curves with points 
  show the integral distributions for the minimum and maximum
  masses of the luminous stars with SED models from Table~\protect\ref{tab:stars}.  
  The maximum mass star is $\gamma^2$~Vel.  The maximum (minimum) masses
  correspond to younger (older) ages, and the CMD models strongly
  favor an older population and so match the minimum mass
  curve well.
  }
\label{fig:mdist}
\end{figure}

\begin{figure}
\includegraphics[width=0.50\textwidth]{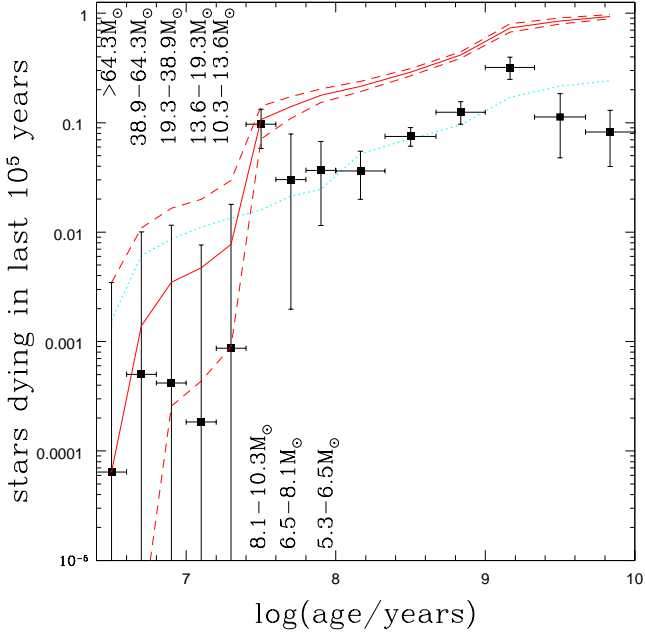}
\caption{
  The number of stars dying in the last $10^5$ years
  for each age bin.
  The points show the probability for each bin and its
  90\% confidence range.  The horizontal error bars span
  the bin widths, and the mass range corresponding to each
  age bin is listed.  The solid red curves shows
  the median integral distribution and the dashed curves
  show its 90\% confidence range.  The arbitrarily normalized
  dotted cyan curve shows what the age distribution would
  be for a constant star formation rate. 
  All uncertainties are
  highly correlated.    
  }
\label{fig:sn}
\end{figure}

\begin{figure}
\includegraphics[width=0.50\textwidth]{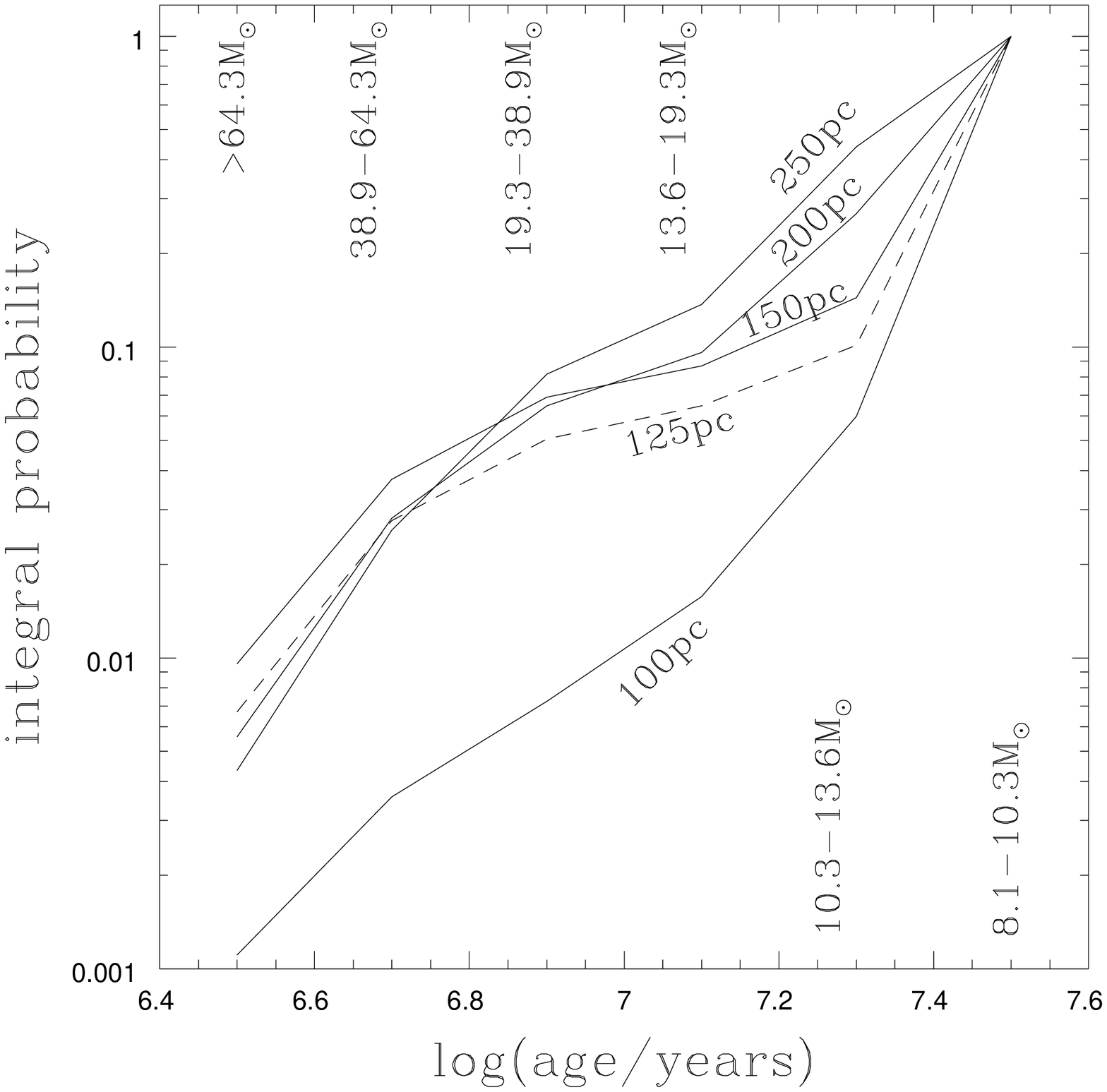}
\caption{
  Integral probability distributions for the age of the progenitor
  of Vela as a function of the radius $R$ used to select the
  stars.  The dashed curve is for the $R=125$~pc radius used
  until now.  The large difference between this distribution 
  and the distribution for $R=100$~pc is due to the inclusion
  of $\gamma^2$~Vel in the larger sphere.
  }
\label{fig:sn3}
\end{figure}

\section{The Progenitor of Vela}

The striking property of Figures~\ref{fig:cmd} and the models
in Table~\ref{tab:stars}
is the marked absence of very young stars or massive stars with 
the exception of $\gamma^2$~Vel. All of the other stars are
only consistent with populations older than ten million years,
corresponding to maximum progenitor masses $\ltorder 15M_\odot$.  The
next step is to use these stars to estimate the mass of the 
progenitor of the Vela pulsar.  Here we will model the Gaia
color-magnitude distribution to estimate the numbers of stars
as a function of age.  As discussed in \S2, we present all of
the results for a $125$~pc sphere around Vela, and the estimated
age/mass of the progenitor for regions from $100$ to $200$~pc in radius.  

We will assume a \cite{Salpeter1955} initial mass function (IMF) with
a minimum mass of $M_{min}=1M_\odot$ since we are not interested in 
the lower mass stars.  If the star formation rate ($M_\odot$/year) 
for $M > M_{min}$ stars is $SFR$, then the rate of forming stars is
\begin{equation}
     { d N \over dt d M } = { (x-2) SFR \over M_{min}^2 }
            \left( { M \over M_{min} }\right)^{-x}
\end{equation}
with $x=2.35$ and a mean mass of $\langle M \rangle = (x-1)M_{min}/(x-2)$.
We divide the star formation history into logarithmic time intervals
$t_{min,i} < t < t_{max,i}$ with $\Delta t_i = t_{max,i}-t_{min,i}$.
Assuming the star formation rate $SFR_i$ in the interval is constant,
the number of $M> M_{min}$ stars formed is 
\begin{equation}
     N_i = { SFR_i \Delta t_i \over \langle M \rangle }.
\end{equation} 
The number of stars formed in this period which die in a short time 
interval $\delta t$ today is
\begin{equation}
       N_i { \delta t \over \Delta t_i } 
        \left[ \left( { M(t_{min,i}) \over M_{min} }\right)^{1-x}
         - \left({ M(t_{max,i}) \over M_{min} }\right)^{1-x} \right]
          = N_i S_i \delta t
       \label{eqn:ndie}
\end{equation}
where $M(t)$ is the most massive surviving star on the isochrone, and
$S_i \delta t $ is the fraction of $M>M_{min}$ stars that died in the last
$\delta t$ years.
This expression is explicitly assuming single star evolution.
We used $8$ temporal bins spanning $6.4 < \log t < 8.0$, bin widths 
of $0.2$~dex and use the number $N_i$ of $M>M_{min}$ stars formed 
as the variable to be determined.  

If the global IMF is \cite{Salpeter1955} down to $0.5M_\odot$ and
then flattens to $M^{-1.3}$ from $0.08 M_\odot$ to $0.5M_\odot$
(\citealt{Kroupa2001}), then the $M>M_\odot$ stars represent
9.1\% of the stars formed.  The global mean mass is $0.61M_\odot$,
so for each $M>M_\odot$ star formed, the total mass of new stars 
is $6.7M_\odot$. Local estimates of the mean stellar mass density
are $0.04 M_\odot$~pc$^{-3}$ (e.g., \citealt{Flynn2006}), so we
should find $\sum N_i \simeq 6 \times 10^4$.  
If we use $2\pi R_d^2 H$ with $R_d=3$~kpc
and $H=0.1$~kpc as the volume of the Galaxy, forming ten $M>M_\odot$
stars per megayear in a 125~pc sphere corresponds to
a global star formation rate of $\sim 0.04 M_\odot$/year. 

We start by building density maps
$\rho_i^{jk}$ in absolute magnitude $M_G$ (index $j$)
and color $B_P-R_P$ (index $k$) for each time interval $i$.  We use Solar 
metallicity {\tt PARSEC} isochrones sampled at $\Delta \log t = 0.01$~dex.
For each density map $i$ we carry out $N_{trial}$ trials.  We 
uniformly select a time between $t_{min,i}$
and $t_{max,i}$, which corresponds to assuming a constant star formation
rate for each bin,  randomly draw an initial stellar mass from
the IMF, and use the isochrones to determine
the absolute magnitude and color if a star of this mass still
exists. 

For the magnitude range we consider, the uncertainties in the
Gaia magnitudes are negligible.  Similarly, the median parallax
uncertainty of $0.027$~mas is also unimportant given that the
smallest parallax we consider is $2.43$~mas. The median
{\tt mwdust} extinctions increases as 
$E(B-V) = 0.19 (d/\hbox{kpc})$ with distance $d$ from the Sun.  The width
of the distribution is approximately $0.13(d/\hbox{kpc})$.
Since the {\tt mwdust} estimates agreed reasonably well with
the extinction estimates from the SED fits, we randomly assigned
each trial the {\tt mwdust} extinction of one of the $M_G<0$~mag
stars.

\def\maglim{0.0}
If the extincted trial star has $M_G < \maglim$~mag, we add $N_{trial}^{-1}$ to the 
appropriate cell of $\rho_i^{jk}$.  The maps span a finite range
of color ($-0.75 < B_P-R_P < 3.5$) and absolute magnitude 
($\maglim > M_G > -8.0$) so trial stars that fall outside the color
range or are brighter than $-8.0$~mag are placed at the bin edge. 
This preserves the total probability for $M_G < \maglim$~mag stars.
As we build the maps, we also build a binned mass distribution
of the $M_G < \maglim$~mag stars, $D_i^j$ for mass bin $j$.  
For $M_G < \maglim$~mag we are including almost all $M>5M_\odot$
stars, and then only mass-dependent portions of the post-main
sequence lifetimes of the lower mass stars. 

With these definitions, the number distribution of stars in magnitude 
and color brighter than the selection limit from time period $i$
is $ N_i \rho^i_{jk}$ and $N_i \sum_{jk} \rho_i^{jk} = N_i F_i$ is the number of 
stars still living and passing the selection criterion.  $F_i$ is
the fraction of $M>M_{min}$ stars which are more luminous than
$M_g > 0$~mag.
The expected number of stars in a color/absolute magnitude bin is 
\begin{equation}
      e_{jk} = \sum_i  N_i \rho_i^{jk}.
     \label{eqn:num1}
\end{equation} 
If we now distribute the $N$ actual stars over the grid,
the observed number of stars in a cell is $n_{jk}$ with a
Poisson probability of finding that number of stars of
$e_{jk}^{n_{jk}}\exp(-e_{jk})/n_{jk}!$.  The logarithm of
the likelihood for all $N$ stars is 
\begin{equation}
    \ln L = \sum_{stars} \ln \left({ r e_{ij}^{n_{ij}} \over n_{ij}! }\right)
     - \sum_{all} r e_{ij}
     \label{eqn:like1}
\end{equation}
where the first term is the sum over bins containing stars
and the second is the sum over all bins.  Note that the
factorial $n_{ij}!$ can be discarded since the calculation
depends only on likelihood differences and not the 
absolute likelihood.  Putting trial
stars falling off the grid when constructing the 
distributions on the bin edges ensures
that the second sum is correct.

In Equation~\ref{eqn:like1} we have also introduced a ``renormalization'' 
factor $r$.  Equation~\ref{eqn:like1} with $r=1$ still includes
the Poisson uncertainties from the total number N of stars being
modeled.  For the
problem of estimating the progenitor mass, however, we 
need the relative probabilities (i.e. ratios) of the numbers of stars in 
the age bins, not their absolute values, and the ratios are 
unaffected by Poisson fluctuations from the finite number of stars.
We solve this problem by optimizing the likelihood with 
respect to the renormalization factor to find that
\begin{equation}
       r = N \left[ \sum_{all} e_{jk} \right]^{-1},
      \label{eqn:renorm}
\end{equation}
which we then use to renormalize $N_i \rightarrow rN_i$.
With this renormalization, $\sum_{all} e_{jk} \equiv N$ and
we have effectively converted the Poisson likelihood into a
multinomial likelihood for how to divide the $N$ stars over
the age bins.  

Operationally, we optimize the likelihood and estimate the
uncertainties using Markov Chain Monte Carlo (MCMC) methods with
the $\log N_i$ as the variables.  Once trial values of $\log N_i$
are selected, they are renormalized before computing the
likelihood.  Some age bins were susceptible to 
$\log N_i \rightarrow -\infty$ (i.e., $N_i$ arbitrarily
close to zero), so we added a weak
prior that the star formation rates of adjacent temporal
bins should be similar by adding
\begin{equation} 
    \lambda^{-2} 
      \sum_i
   \left[ \ln \left( {N_i \Delta t_{i+1} \over N_{i+1} \Delta t_i }\right)\right]^2
\end{equation}
with $\lambda = 6.91$ to the likelihood.  This adds a penalty
of unity to the likelihood if adjacent bins have star formation
rates ($SFR_i \propto N_i/\Delta t_i$) that differ by a factor
of 1000.  This is just to prevent 
numerical divergences and has no significant effect on the results.  

Figure~\ref{fig:cmd} shows contours for the numbers of
stars as a function of luminosity and temperature for
the maximum likelihood model. Not surprisingly the 
maximum density lies along the main sequence of B
stars.  The density contours follow the stellar distribution
closely.  There are a few red giants in very low density
regions.  This may mean that there some stars with higher
extinctions than our model, probably cases which required
some circumstellar dust.  They are, however, old
lower mass stars and should have no effect on the inferences
about young, high mass stars.

Figures~\ref{fig:age}, \ref{fig:age2} and \ref{fig:sfr}
show three different ways of viewing the distribution of
stars in age.  Figure~\ref{fig:age} shows the distribution
of the modeled stars, $N_i F_i$, in age. Because of the
renormalization procedure, the numbers add up exactly to
the $N=607$ stars used. The five youngest age bins, 
corresponding to progenitors more massive than $10.3M_\odot$
contain a median of $N_iF_i=5$  modeled stars and fewer than 19 at
95\% confidence.  For these young bins, we
are basically counting all $M \gtorder 5M_\odot$ stars,
so the number of higher mass stars is lower.  For ten
$M > 5 M_\odot$ stars and a \cite{Salpeter1955} IMF,
there are $3.9$, $2.3$, $1.5$ and $0.9$ stars more 
massive than $10$, $15$, $20$ and $30M_\odot$, respectively.
The field contains very few high mass stars, 
consistent with Figure~\ref{fig:cmd}.  
For comparison, the $10^{7.4}$-$10^{7.6}$~year bin, which spans a 
final mass range of $8.1$ to $10.3M_\odot$
that corresponds to the minimum mass range
expected to produce supernovae for single stars, has
a median of $N_i F_i=95$ modeled stars.

Only a small fraction of the $M>M_{min}$ stars are luminous
enough to be selected with $M_G>0$~mag, so the actual number
of stars associated with  the older bins is much larger.  Figure~\ref{fig:age2}
show the number $N_i$ of $M>M_{min}$ stars as a 
function of age.  As expected, the total number of 
$M>M_\odot$ stars ($(6.8\pm 2.3) \times 10^4$)
in the volume is much larger than the $607$ we modeled.
and the total number of $M>0.08 M_\odot$
stars would be another $ \sim 11$ times larger. The 
implied stellar mass density of $(0.056 \pm 0.019)M_\odot$~pc$^{-3}$
is remarkably close to the local estimate of 
$0.05 M_\odot$~pc$^{-3}$ (\citealt{Flynn2006}) given the huge correction 
required to go from the number of $M_G>0$~mag stars to the 
total mass of stars - for the oldest
bin, the $\sim 50$ modeled $M_G>0$~mag stars represent some $\sim 20,000$
$M>M_\odot$ stars which then correspond to some $\sim 200,000$
stars in total.  

Finally, Figure~\ref{fig:sfr} shows the star formation rates
for each temporal bin as the number of $M>M_\odot$ stars formed
per $10^6$~years.  As discussed above, a formation rate of 10 $M>M_\odot$
stars per million years corresponds to a global Galactic star
formation rate of order $0.04 M_\odot$/year, so the peak 
star formation rate of $\sim 0.3 M_\odot$/year on a global 
basis is not impressive.  The peak is, however, again found
in the $10^{7.4}$-$10^{7.6}$~year bin.  The star formation
rates in the younger bins are at least an order of magnitude
lower, with even fewer new stars because the associated times
$\Delta t_i$ become shorter (i.e. the drop off in 
Figure~\ref{fig:age2} is sharper than in Figure~\ref{fig:sfr}).

Figure~\ref{fig:mdist} shows the model for the integral 
distribution of the selected stars in mass
\begin{equation}
     D^j = \sum_i N_i D_i^j
\end{equation}
where the $D_i^j$ are the mass distributions associated
with each temporal bin.
As expected from the distributions in Figures~\ref{fig:cmd}, 
the region contains few higher mass stars.  
We also show the integral
distribution of the mass estimates
from the SED models in Figure~\ref{fig:mdist}.  Broadly speaking they are
in good agreement, although the results from 
the CMD models track the
lower mass limits much more closely than the upper mass
limits.  As noted earlier, at roughly constant luminosity and temperature,
there is a strong correlation between age and mass for
these main sequence B stars in the sense that 
the minimum mass estimates correspond to the maximum
ages and {\it vice versa}.  The CMD models strongly
disfavor having significant numbers of stars with
the young ages associated with the maximum mass 
estimates and strongly favor ages corresponding to the 
minimum mass estimates and so track the minimum mass 
estimates.  The CMD models constrain the age distribution
better than the SED models of individual stars.

The outlier in Figure~\ref{fig:mdist} is $\gamma^2$~Vel.  
The model mass distribution based on the CMD
predicts only $\simeq 0.3$ ($\simeq 0.1$) stars with 
$M>20M_\odot$ ($>30M_\odot$).  The existence of 
$\gamma^2$~Vel is not a huge statistical anomaly, 
as the likelihood of having one or more such stars
is $26\%$ ($10\%$).  However, in the CMD (Figure~\ref{fig:cmd}), 
there are simply no other stars between where the 
$10^{7.5}$~year isochrone (maximum mass $9.1M_\odot$) 
starts to turn off the main sequence 
and $\gamma^2$~Vel. The absence of these stars drives
the model to make $\gamma^2$~Vel moderately unlikely.

Figure~\ref{fig:sn} shows the number of stars $N_i S_i \delta t$
(Equation~\ref{eqn:ndie}) predicted
to have died in the last $\delta t = 10^5$~years for each
age bin. That the integral probability sums to near unity
is happenstance -- the distribution is not normalized.  
The probability for the ages leading to
supernovae is dominated by the 
age range corresponding to initial masses of $8.1$-$10.3M_\odot$.
Note that for the range of ages producing supernovae,
the expected number of deaths in the last $10^5$~years
is only $\sim 0.1$ and so
the time scale for this volume to produce
a supernovae is $\delta t \simeq 10^6$~years, far longer
than the life time of an SNR.  This is not a statistical
problem -- we chose this volume because it contained
an SNR.  To estimate $\delta t$ we would have to analyze 
a much larger volume chosen without using any prior knowledge
of the number of enclosed remnants. 

Figure~\ref{fig:sn} also shows what the prediction would
be if the star formation rate was constant across all age
bins, roughly normalized to the intermediate age bins where
the estimated star formation rates are nearly
constant (Figure~\ref{fig:sfr}).  The four youngest and two
oldest bins have lower estimated SFRs, and so fewer stars
are predicted to have died in the last $10^5$~years, while 
the $10^{7.4}$-$10^{7.6}$~year bin has a higher SFR and
more predicted deaths.  The difference between the two
curves represents the information added by using the local
stellar population to infer the star formation history. 

The probability of a star dying in the last $\delta t = 10^5$
years is dominated by the older bins and the volume is likely
to contain a number of young white dwarfs.  Out of curiosity
we did a cursory search for candidates.  Based on the MIST
(\citealt{Dotter2016}, \citealt{Choi2016}) evolutionary tracks, a $<10^5$~year old white dwarf
should be a very hot star with a luminosity $\sim 10^2 L_\odot$.    The Gaia EDR3 
source ID\#5319832121597913984 is the bluest and most luminous
source near the tip of the usual white dwarf cooling 
sequence.  It is flagged as a very high probability 
white dwarf candidate by \cite{Gentile2019} but with no
estimates of its physical properties. There are no
UV or spectroscopic observations of it, but for an
assumed temperature of $40,000$~K it would have a 
luminosity of $\sim 2 L_\odot$.  Since the luminosity
scales linearly with the temperature on the Rayleigh-Jeans
tail of the SED, this star has to be significantly
older than
$10^5$~years even at twice the temperature.  There are five similarly blue
but $\Delta M_G \sim -3$~mag more luminous stars (\#5592257426113535104,
\#5411002594979483392, \#5442024044243183232, \#5512856125193586176, and
\#5539780553619327360) whose luminosities could be in
the right range. Three of them are spectroscopically classified
as hot sub-dwarfs (\#5592257426113535104, \cite{Garrison1973};
\#5442024044243183232, \citealt{Barlow2013}; \#5539780553619327360,
\citealt{Kilkenny1988}, one (\#5411002594979483392) 
is spectroscopically classified as an A0 star (\citealt{Nesterov1995},
but this difficult to reconcile with its location in the CMD),
and \#5512856125193586176 has no spectroscopic classifications.
Possibly one of these latter two sources is a very young white
dwarf masquerading as a hot subdwarf.

To produce a final constraint on the mass of Vela's progenitor we
must consider three remaining issues. First, we have to put a 
minimum mass in by hand.
When analyzing a single region around one target, we cannot determine
a minimum mass for explosion.  This requires analyzing many such regions
both with and without SNRs.  We must
impose a minimum mass by simply dropping the older age bins.  The obvious
choice is to keep only the $10^{7.4}$-$10^{7.6}$~year, with its
maximum masses at death of $8.1$-$10.3M_\odot$, and younger bins.
For single star evolution, this is the correct mass range for the
cutoff.  With binary evolution the next older bin can contribute
through mergers of longer lived, lower mass stars (see, e.g., \citealt{Zapartas2017}).
Retaining this next older bin would only strengthen our final
conclusion that the progenitor was significantly less massive than $20M_\odot$. 
The change would be modest because many fewer stellar deaths 
are expected from this age bin than from the $10^{7.4}$-$10^{7.6}$~year 
bin (see Figure~\ref{fig:sn}).

In our formalism, the number of stellar deaths is $N_i S_i \delta t$
where $S_i$ is independent of $\delta t$. For our final result we
want the relative probabilities of the age bins with no dependence
on $\delta t$.  The probability
that the progenitor came from bin $i$ and no other bin $j$ is
\begin{eqnarray}
   P(i \& !j) &=   &N_i S_i \delta t \exp(-N_i S_i \delta t)
     \Pi_{i\neq j} \exp(-N_j S_j \delta t) \\
     &= &N_i S_i \delta t \Pi_{all} \exp(-N_J s_j \delta_j). \nonumber
\end{eqnarray}
The total probability summed over all the bins is
\begin{equation}
  P_{tot} = \sum_i P(i \& !j) = \delta t \left[\sum_{all} N_i S_i\right]
        \exp\left( - \delta t \sum_{all} N_i S_i \right)
\end{equation}
and thus the normalized probability for each bin of
\begin{equation}
   P_{tot}^{-1} P(i \& !j) =
   N_i S_i \left[ \sum_{all} N_i S_i \right]^{-1}
\end{equation}
is independent of $\delta t$ as desired.  Basically, this just
corresponds to normalizing the integral probability distribution
so that the total probability is unity. 

Finally, we have a maximum likelihood solution and all the MCMC
samples, each of which represents a realization of the integral
probability distribution.  We could go to each age bin, sort 
these distributions and report a median and some range, say 90\%
confidence, but it is unclear how to interpret this.  What does
it mean to say there is a 5\% chance that there is a 10\% chance
that the progenitor was younger than the age of some bin?  There
really should be only one distribution which incorporates this information.
We are really combining the $N_{MCMC}$ results of the MCMC chains, each 
of which has a probability of $N_{MCMC}^{-1}$, and each trial predicts a 
probability for the age of the progenitor.  The way to combine
them is to sum the probability the trial predicts for the age
bin weighted by the probability of the trial -- in short, the
final probability distribution is simply the average of the MCMC
samples.

Figure~\ref{fig:sn3} shows these final probability distributions
as a function of the radius $R$ of the sphere used to select the
stars. Recall that the results so far have all been for $R<125$~pc.
The maximum likelihood age distributions modestly favor
older, lower mass progenitors compared to these distributions. 
For $R < 100$~pc, a low mass progenitor is very strongly favored
with a 94\% probability of it coming from the lowest mass bin
($8.1$-$10.3M_\odot$).  However, as we discussed in \S2, this
sphere probably only contains $\sim 2/3$ of the stars born 
within a similar radius.  If we increase the radius to $R<125$~pc,
the structure of the distribution changes considerably, and
this is entirely driven by the inclusion of $\gamma^2$~Vel.  The
probability of the oldest age bin is still high (90\%), but
the probability of an age bin corresponding to a progenitor
more massive than $>20M_\odot$ increases by almost an order of
magnitude.  None the less, the probability of a progenitor less
massive than $20 M_\odot$ is still 95\%.  Expanding the sphere
still further primarily increases the probability of the 
$13.6$-$19.3M_\odot$ and $10.3$-$13.6M_\odot$ progenitor bins
relative to the $8.1$-$10.3M_\odot$ bin.  The probability of
being less massive than $20M_\odot$ is still $90\%$.   

In fact, $\gamma^2$~Vel is associated with a concentrated
cluster of pre-main sequence stars with a very low velocity
dispersion (e.g., \citealt{Franciosini2018}).  This means
that associating the progenitor of Vela with the formation
of $\gamma^2$~Vel and its cluster requires the progenitor
to have been a runaway star.  To explore this we used the
fiducial runaway model from \cite{Renzo2019} and estimated
the distance the surviving stars from disrupted binaries
could travel in the time left for their current evolutionary
phase.  Since we now have an isolated
star, this will be close to the remaining life time of the
star.  Because of mass transfer, we examined the fraction
of stars which could travel more than $100$~pc (the distance
from $\gamma^2$~Vel to the pulsar) in bins of either the
zero age main sequence (ZAMS) mass or the mass after the 
binary is disrupted.

For the ZAMS mass and no restriction on the mass of the star
which died to disrupt the binary, the chances of traveling
100~pc were significant only for $10$-$15M_\odot$ (20\%)
and $15$-$20M_\odot$ (5\%) stars.  Stars with higher post-disruption
masses could do so because they are initially lower mass
stars that were mass gainers when the primary began to evolve.
Even so, only 7.6\%, 2.7\% and 2.2\% of stars with post
disruption masses of $20$-$25$, $25$-$30$ and $30$-$35M_\odot$
could travel 100~pc.  Demanding a high ($M>30M_\odot$)
ZAMS mass primary helps only modestly. Thus,   
a volume including $\gamma^2$~Vel
does significantly increase the probability of a higher 
mass ($\gtorder 20 M_\odot$) progenitor, but a Vela
progenitor formed in the cluster associated with $\gamma^2$~Vel would
(a) have to be a runaway star from a disrupted binary,
and (b) would have to have a statistically unlikely 
runaway velocity.

\section{Discussion}

The environment of the Vela pulsar is dominated by B stars with
only one nearby O star, $\gamma^2$~Vel.  If we consider only
stars within $R<100$~pc of Vela, the most likely (95\%) age
range for its progenitor is $10^{7.4}$-$10^{7.6}$~years 
corresponding to a mass range of $8.1$-$10.3M_\odot$.  There
is clearly a local burst of star formation associated with
this age bin. For
these ages and the observed velocity dispersions of the stars,
a radius of $R<100$~pc will not encompass all of the stars
formed within 50-100~pc of the progenitor.  Any larger sphere
encompasses the most massive nearby star, the O star plus
WR star binary $\gamma^2$~Vel.  So for $R<125$~pc, there is
still a 90\% probability of associating the progenitor with 
the  $10^{7.4}$-$10^{7.6}$~year age bin, but there is now a 
5\% chance of an age allowing masses $>20 M_\odot$.  However,
$\gamma^2$~Vel is associated with a very low velocity dispersion
cluster, so for the progenitor to be that massive it would also
need to be a runaway star from a disrupted binary with an 
unusually high velocity based on the models of \cite{Renzo2019}.   

We have assumed single star evolution models in this analysis.
We know that Vela itself was not a binary or triple at the time
of its death (\citealt{Kochanek2021}, \citealt{Fraser2019})
down to companion mass limits $\ltorder M_\odot$.  This does
not rule out the progenitor as a merger remnant or as an
unbound secondary from a previous explosion.  As a merger
remnant or an unbound secondary that gained significant mass
through mass transfer, we have underestimated the probability
of a low mass progenitor because we imposed the minimum mass/maximum
age limit by hand since it cannot be determined by analyzing
a single system.  The changes from including the next lower
mass bin would not be huge, because the star formation rate 
estimated for this next bin is much lower than for the
$10^{7.4}$-$10^{7.6}$~year age bin.

As noted in the introduction, the two fundamental limitations
to applying this method in the Galaxy are estimating the 
distance to the
supernova remnant and extinction.  There are three additional
systems which can be analyzed easily.  G180.0$-$01.7 is 
associated with the radio pulsar PSR~J0538$+$2817, which has
a VLBI parallax (\citealt{Ng2007}, \citealt{Chatterjee2009}).
In \cite{Kochanek2021} we examined the properties of the nearby
luminous stars but did not carry out a formal analysis as done
here for Vela.  G205.5$+$0.00.5 (Monoceros Loop) and G284.3$-$01.8
both contain neutron star high mass X-ray binaries 
(\citealt{Hinton2009}, \citealt{Corbet2011}) and the stellar
companions have Gaia parallaxes (see \citealt{Kochanek2021}). 
Unfortunately, the interacting (probably) black hole binary SS~433 
(see the review by \citealt{Margon1984}) in G039.7$-$02.0 is 
likely too distant ($\varpi^{-1}=8.5$~kpc) to use Gaia parallaxes 
to trivially select stars in a $\sim 100$~pc sphere around the binary.
The Gaia EDR3 parallax of SS~433 is also in strong ($\sim 5\sigma$)
disagreement with the distances estimated from kinematic models
of the relativistic jets (\citealt{Blundell2004}, \citealt{Marshall2013}).
As discussed in \cite{Kochanek2021}, the success of \cite{Cha1999}
in determining the distance to the Vela remnant based on the
distance at which stars began to show absorption features from
the supernova remnant provides a simple observational approach
to better determining the distances to other remnants. 

If we consider the 165 pulsars with parallaxes in the 
ATNF Pulsar Catalog (\citealt{Manchester2005}), the vast
majority cannot be analyzed using this method because
the formation region cannot be well-localized. In
particularly, the only available age estimate is the spin 
down age which at best estimates the time of explosion
to within a factor of two.  If we require that the 
projected distance traveled in the spin down age is less than 
$100$~pc, only 10 pulsars are left after excluding Vela.  Because 
the proper motions represent two
components of the kick velocity while the motion along the line
of sight contains only one component, the line of sight motions
should be less of a problem.  If we assume a typical neutron star kick 
velocity of $265$~km/s (\citealt{Hobbs2005}), the typical line-of-sight
distance traveled in the spin down time for the pulsars with transverse
motions less than $100$~pc ranges from 24 to 152~pc.
If we require this line-of-sight motion to be $<100$~kpc and 
restrict the parallax distance to be $<2$~kpc, we are left with
four systems other than Vela (J0157$+$6212, J0633$+$1746, 
J0659$+$1414 and J2032$+$4127).  This would be a sample biased
towards supernovae with low kick velocities.

In the absence of good parallaxes,
the alternative is to simply look at all the stars projected within
a fixed projected separation from the SNR over some broad line
of sight distance range consistent
with estimates for the distance to the SNR.  This is what is done
for all the extragalactic analyses since there is no possibility
of using parallaxes to remove the foreground and background 
contamination.
However, while some of the external galaxies that have been examined are 
highly inclined (e.g., Andromeda at $\sim 13^\circ$), none have
as unfavorable geometry for minimizing contamination as we face
for examining Galactic SNRs. Which Galactic SNRs are suitable for 
such an analysis will depend on their Galactic coordinates.

Related to this is the question of completeness and contamination in
analyses of stellar populations near extragalactic supernovae and 
remnants.
We can use our nearly complete knowledge of the environment around
Vela to evaluate the 50~pc projected search region used by \cite{Jennings2012}
and subsequent papers.  We again restrict ourselves to the more massive
stars using the magnitude and color cuts described in \S2.
We transform the stellar positions to axes
aligned with Galactic coordinates, select stars in a sphere around
Vela, and then count stars using their positions $10^7$~years ago
in circles centered on their median position
as if we were looking down on the plane of the Galaxy
(i.e. we ignore the distance of the star from the Galactic plane).
As before, there are 19 stars in a 50~pc sphere around Vela, while
a 50~pc circle centered on their median position $10^7$ years ago
contains only 9 of them along with 22 other stars, so the completeness
is 47\% and the contamination is 79\%.  A 100~pc circle contains 16
of the stars along with 140 other stars leading to a completeness of
84\% and a contamination rate of 90\%.  If instead we consider a 100~pc
sphere, which contains 152 stars, a 100~pc circle contains 73 of them
with 73 additional stars for a completeness of 58\% and a contamination
rate of 50\%.  For a 150~pc circle, the completeness and contamination 
increase to 76\% and 63\%, respectively.  This age corresponds to
the life time of a $\sim 20M_\odot$ star -- a $\sim 10 M_\odot$
star lives three times longer so the completeness will be far lower
and the contamination will be far higher. For the 50~pc sphere, the
completeness/contamination for the 50~pc and 100~pc circles are
11\%/89\% and 26\%/90\%, respectively. For the 100~pc sphere, 
the completeness/contamination for the 100~pc and 150~pc circles
are 14\%/56\% and 28\%/64\%, respectively.  

If the environment of Vela is typical, a 50~pc projected search radius 
includes almost none of the $\gtorder 5 M_\odot$ stars that were within 50~pc
of the progenitor at birth -- the stars within this radius are 
overwhelmingly stars which were more distant from the progenitor at
birth.  A larger, 100~pc projected search radius does capture a 
reasonable fraction of the stars born within 100~pc of the progenitor,
but they are still only about half of the stars within that projected
radius.  This strongly suggests that this method of analysis cannot
estimate the progenitor properties for individual supernova remnants
or supernovae in external galaxies-- it can only statistically estimate 
the progenitor properties for large ensembles of targets.  Even with
complete three dimensional information, as we have for Vela, there
are significant problems with completeness and contamination in such
small regions on the 30 million year time scale before $\simeq 10M_\odot$ 
stars explode.

\section*{Acknowledgments}

The author thanks K.~Auchettl, J.~Johnson, M.~Pinsonneault, K.~Stanek, and 
T.~Thompson for discussions.
CSK is supported by NSF grants AST-1908570 and AST-1814440.  
This research has made use of the VizieR catalogue access tool, CDS,
 Strasbourg, France (DOI : 10.26093/cds/vizier). The original description 
 of the VizieR service was published in \cite{Ochsenbein2000}.
This research has made use of the SIMBAD (\citealt{Wenger2000}) database,
operated at CDS, Strasbourg, France.

\section*{Data Availability Statement}

All data used in this paper are publicly available.

\end{document}